\documentclass[prl,aps,twocolumn,amsmath,amssymb,floatfix]{revtex4}
\usepackage{booktabs,graphicx,verbatim}
\usepackage[colorlinks=true,citecolor=blue,filecolor=blue,linkcolor=blue,urlcolor=blue,pdftex]{hyperref}
\usepackage{ulem}
\begin{document}
\title{Reply to the Comments on the XENON100 first Dark Matter Results}
\author{The XENON100 Collaboration}\noaffiliation
\maketitle

The recently submitted preprint on the first results from the XENON100 dark matter experiment \cite{ref:xe100} was followed by a criticism by J.I.\:Collar and \mbox{D.N.\:McKinsey}  \cite{ref:collar}, focused on our extrapolation of the scintillation efficiency ${\cal L}_{\text{eff}}$ to the lowest nuclear recoil energies, where no data and no theoretical model exist. Here we 
add clarifications on our analysis and comment on their criticism.

The main XENON100 result demonstrates, with only 11 days of data, the potential of this ultra-low background experiment to exclude new parameter space in spin-independent WIMP-nucleon cross section. The high mass WIMP limit is unaffected by the controversy on the value of ${\cal L}_{\text{eff}}$.
The light WIMP interpretation of the DAMA \cite{ref:dama} and CoGeNT \cite{ref:cogent} data is excluded at 90\% CL for a constant extrapolation of the global fit of all  ${\cal L}_{\text{eff}}$ data from fixed-energy neutron experiments (see Fig.\:1 of our preprint). However, when  ${\cal L}_{\text{eff}}$ is assumed to follow the 90\% lower limit of the global fit with a conservative logarithmic extrapolation to zero near 1~keV$_r$, a fraction of the CoGeNT parameter space remains uncovered. 

We further argue that we don't see any events even down to a threshold of 3 PE, departing from our a priori chosen threshold of 4 PE, where, however, our acceptance is still quite high (as shown in Fig.\:2 of the preprint). This excludes all of the DAMA favored region, with ion channeling, at 90\% CL, even for the conservative case of ${\cal L}_{\text{eff}}$. In this case, we would expect one event for a 7 GeV/c$^2$ WIMP, at the lower edge of the CoGeNT favored region. With our current data, only this conservative ${\cal L}_{\text{eff}}$ leaves room for low mass WIMPs compatible with CoGeNT in the mass range of $\sim$7--9 GeV/c$^2$.

We will now comment on some of the statements made in \cite{ref:collar} in more detail.

\begin{figure}[htb]
\includegraphics*[width=8.5cm]{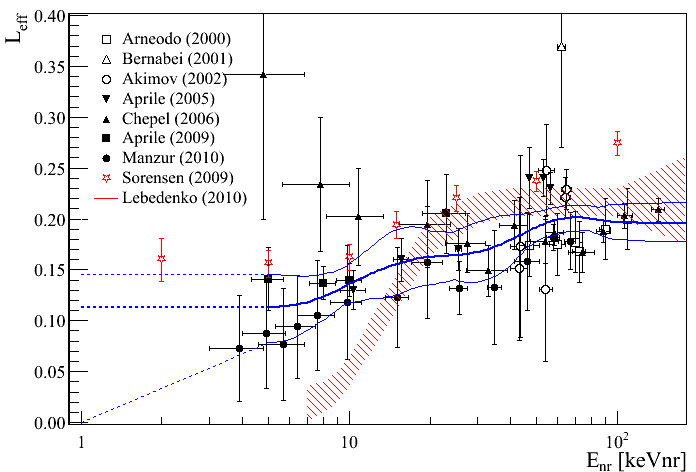}
\caption{All published data on ${\cal L}_{\text{eff}}$: The black datapoints -- used for the global fit in the XENON100 paper \cite{ref:xe100} -- are all published direct measurements of ${\cal L}_{\text{eff}}$. The red data (Sorensen (XENON10) \cite{ref:xe10Leff} and Lebedenko (ZEPLINIII) \cite{ref:zep3}) are from comparisons of data with Monte Carlo simulations. They were not
used on the global fit because of their possibly larger systematic uncertainties. The three blue solid contours are the result from a global fit to all direct measurements (black) in the region from 5 -- 100~keV$_r$. The thinner contours above and below are the $\pm$90\% confidence level contours. The dashed lines below 5~keV$_r$ are the extrapolations as explained in the text. For the first XENON100 data analysis, only the best fit and the lower 90\% CL contour are used.} \label{fig:Leff}
\end{figure}

1. In \cite{ref:collar} it is stated that we use a constant  ``${\cal L}_{\text{eff}}\!\sim\!0.12$ below $\sim\!10$~keV$_r$''. This statement is not correct. As explained and shown in Fig.\:1 in the preprint, and also shown in Fig.\:\ref{fig:Leff} here, the ${\cal L}_{\text{eff}}$ used in our analysis is not based on a single measurement (given the apparent disagreement within the quoted errors) but is the result from a global fit to all existing direct neutron scattering measurements. 
These results, where the recoil energy is measured by the scattering angle of mono-energetic neutrons, are systematically cleaner than results that are based solely on a comparison between neutron calibration data (featuring single nuclear recoils from a continuous neutron source) and Monte Carlo simulations. This is the reason why we have chosen not to include these ${\cal L}_{\text{eff}}$ results in the global fit, and we have clarified this in the preprint.  For illustration, we have added them in Fig.\:1 of this note. There are two studies of this type (while only one is mentioned in \cite{ref:collar}). The first is from the XENON10 collaboration\:\cite{ref:xe10Leff} and lies above our global fit. It is consistent with the 90\% upper contour of  ${\cal L}_{\text{eff}}$ and clearly supports a constant extrapolation of ${\cal L}_{\text{eff}}$ to lower energies. The ZEPLINIII results \cite{ref:zep3} do not agree with any of the tagged-neutron measurements, but we note that the assumed field quenching at the high operating field of that experiment (3.9 kV/cm) has not been determined experimentally. 

Collar \& McKinsey also point to a recent workshop presentation by a former Ph.D. student of the XENON10 collaboration (their ref. [8]) as ``the most recent'' analysis by the XENON10 collaboration. This result has not even been reviewed within the XENON10 collaboration and is unpublished. If this adds anything to this discussion, it underscores the point that the comparison of simulations and wide-spectrum neutron calibration data is subject to large systematic uncertainties. 
Furthermore, the lowest data points from the referenced slides were omitted in \cite{ref:collar}. These points show a trend which is less supportive of their argument.

We thus hold that at this stage an unbiased treatment of all the direct scattering data sets is the best approach and perform a global fit over all these data for the analysis of XENON100 data.This global fit ${\cal L}_{\text{eff}}$ is monotonically falling with energy and its systematic uncertainty is given by the $\pm$90\% confidence level contours. By using the correct statistical approach of a global fit with corresponding uncertainty, we have chosen not to favor our own ${\cal L}_{\text{eff}}$ measurements over others. On the other hand,  Collar \& McKinsey suggest the use of one of the authors' own ${\cal L}_{\text{eff}}$ measurement \cite{ref:manzur}. We do not agree with their statement that this ``is featuring the best control of systematics so far''. The trigger efficiency at the lowest energies probed in \cite{ref:manzur} is very low, and is not measured but inferred from a Monte Carlo simulation.  The way this rapidly falling efficiency is corrected impacts substantially the derived ${\cal L}_{\text{eff}}$.  In the measurement of Aprile et al.~\cite{ref:aprileLeff}, the trigger efficiency was $\ge$90\% at 5~keV$_r$, and this was directly measured and verified with a simulation. This means that all data points were obtained with similar systematics in \cite{ref:aprileLeff}, a statement that does not hold for \cite{ref:manzur}. The systematic uncertainty from the  spread in the scatter angle due to the finite size of the detectors is also larger in \cite{ref:manzur}. 

2. In their footnote, the authors express their confusion about the ${\cal L}_{\text{eff}}$ curves used in the analysis and accuse us of trying to mislead the reader. We reject this notion. 
To avoid any misunderstandings, we have changed the text slightly to be more clear:
In Fig.\:1 of \cite{ref:xe100} (also in Fig.\:\ref{fig:Leff} here) there are 3 ${\cal L}_{\text{eff}}$ contours: The thicker one in the center is the result from the global fit with a constant extrapolation below 5~keV$_r$ as explained above. The thinner contours above and below are the $\pm$90\% confidence level contours from this fit. To be very conservative, the lower contour is logarithmically extrapolated to energies below 5~keV$_r$, with ${\cal L}_{\text{eff}}=0$ around 1~keV$_r$. The slope of the extrapolation is far from ``arbitrary'' but fixed by a fit to the low energy part of the Yale points \cite{ref:manzur} and matched to the lower 90\% confidence contour at 5~keV$_r$. 

The logarithmic extrapolation is very conservative since a  linear extrapolation describes the low energy part of the data points from ref.\:\cite{ref:manzur} equally well, and would result in a much higher ${\cal L}_{\text{eff}}$ and hence stronger constraints on low-mass WIMPs. 
From the three contours in Fig.~\ref{fig:Leff}, only the central (``global fit'') and the lower one (``lower 90\% CL contour'') are used in the XENON100 analysis, as clearly stated in \cite{ref:xe100}.  

3.  No satisfactory theory describing the behaviour of ${\cal L}_{\text{eff}}$ in liquid xenon exists so far.  The authors state that a kinematic cutoff to the production of scintillation is expected  whenever the minimum excitation energy $E_{\text{g}}$ of the system exceeds the maximum possible energy transfer  to an electron by a slow-moving recoil ion, $E_{\text{max}}$. They refer to papers by Ahlen\&Tarle~\cite{ref:ahlen} and Ficenec, Ahlen, Tarle et al.~\cite{ref:ficenec}. These papers deal with  protons in organic scintillators. Their arguments do not necessarily  apply to Xe-Xe collisions. 
It is known in fact that Lindhard theory\:\cite{ref:lindhard,ref:mangiarotti} is not adequate at very low energies, where mostly the tails of the ion-ion potential are probed and the Thomas-Fermi treatment becomes a crude approximation. For Xe-Xe collisions this corresponds to about 10~keV$_r$. The electron cannot be treated separately from the Xe atom and the maximum energy  transferred to an electron cannot be given by simple kinematics, as advocated in \cite{ref:collar}. 

The collision mechanism for heavy ions at very low energies may be better described by, e.g., the molecular orbit theory \cite{ref:mott}, which involves many-body kinematics. The  argument by Collar and McKinsey is based on two-body kinematics and would not apply for heavy ion collisions in the energy region  concerned here. In fact, Ficenec et al.~\cite{ref:ficenec} state that ``No evidence  for a response cutoff is observed at velocities extending well below the electron-excitation threshold of $6\times10^{-4} c$ expected from two-body kinematics'' even for protons. Besides, if $E_{\text{max}}$ for Xe-Xe is 39~keV$_r$, the kinematics  argument cannot explain the scintillation observed below 39~keV$_r$ at all. Apart from the uncertainty in stopping power calculations which affect directly nuclear quenching, other factors may affect ${\cal L}_{\text{eff}}$ through electronic quenching. However, the current experimental and theoretical situation is such that there is no proven mechanism which justifies a decreasing ${\cal L}_{\text{eff}}$ with decreasing energy, as strongly advocated by Collar and McKinsey. 

We are fully aware of the impact of ${\cal L}_{\text{eff}}$ on the overall sensitivity of noble liquid dark matter experiments and our answer is simply that we will measure it again, extending it to the lowest possible energies. We need accurate data on this quantity and, within the XENON collaboration, we have already developed two new and independent  set-ups optimized to measure the energy and field dependence of both electron and nuclear recoils in liquid xenon.  

\begin{figure}[htb]
\includegraphics*[width=8.5cm]{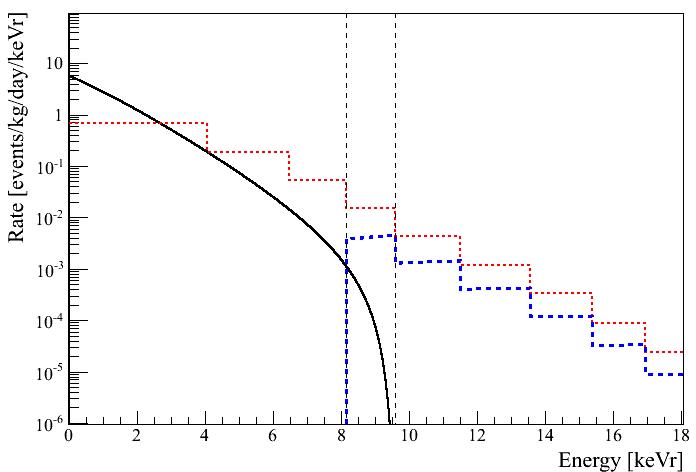}
\caption{Expected spectrum of a 10~GeV/$c^2$ WIMP with a cross section of $1 \times 10^{-41}$~cm$^2$ (black, solid), a benchmark case at the lower edge of the DAMA region. The red (dashed) lines show the spectrum after a convolution with a Poisson distribution, the blue (thick dashed) line is corrected for the XENON100 efficiency. The straight lines are the 3 PE and 4 PE thresholds using the lower 90\% CL ${\cal L}_{\text{eff}}$ contour of the global fit as explained in the text.}\label{fig:spectrum:DAMA}
\end{figure}

4. Finally, Collar and McKinsey doubt that we have properly taken into account the effects of the low number of photoelectrons at our threshold. (Note that this effect had not been accounted for in the preliminary plots presented in their reference [17].) We agree that this has a crucial impact on the XENON100 sensitivity to low mass WIMPs, however, it is a fact that an imperfect threshold due to a finite energy resolution leads to a mixing of events below threshold into the sample and vice versa. Since the expected WIMP spectrum is a steeply falling exponential (see~Fig.\:\ref{fig:spectrum:DAMA}), many more sub-threshold events fall in the energy region above threshold than vice versa.

Due to the low number of detected photoelectrons at the XENON100 threshold, the energy resolution is completely dominated by counting statistics, therefore the expected true differential rate is convoluted with a Poisson function to account for this behavior. We also point out
that the XENON100 efficiency is still very high down to 3\:PE.

Figure\:\ref{fig:spectrum:DAMA} shows the effect of Poisson broadening of our threshold for a DAMA benchmark case: 
There is a small amount of rate from a 10~GeV/$c^2$ WIMP with a cross section of $1 \times 10^{-41}$~cm$^2$ leaking into the XENON100 signal region, even at a threshold of 4~PE, corresponding to 9.6~keV$_r$ in the case of the lower 90\% CL ${\cal L}_{\text{eff}}$ contour. Based on the light WIMP interpretation of the DAMA annual modulation signal,
we would expect to see a total of  4.4~events above a lowered threshold of 3~PE (8.2~keV$_r$ for the conservative ${\cal L}_{\text{eff}}$ case), taking into account our reduced detection efficiency at lower energies (more than 18\:events are expected for the best fit ${\cal L}_{\text{eff}}$ above 3~PE corresponding to 7.0~keV$_r$). As explained in our manuscript (see Fig.~3 in \cite{ref:xe100}) no event is observed leading to an exclusion of this case at 90\% CL.

The same effect at the lower end of the WIMP mass range favored by CoGeNT  (7~GeV/$c^2$ WIMP, cross sections of $0.5$ -- $1 \times 10^{-40}$~cm$^2$) reduces the expectation to 0.73 -- 1.5 events above a threshold of 3 PE in the case of the lower 90\% CL ${\cal L}_{\text{eff}}$ contour. Hence no significant conflict is found under this assumption and with the limited exposure used so far. For the global fit with constant extrapolation of ${\cal L}_{\text{eff}}$, this region is excluded at 90\% CL.

\begin{figure}[htb]
\includegraphics*[width=8.5cm]{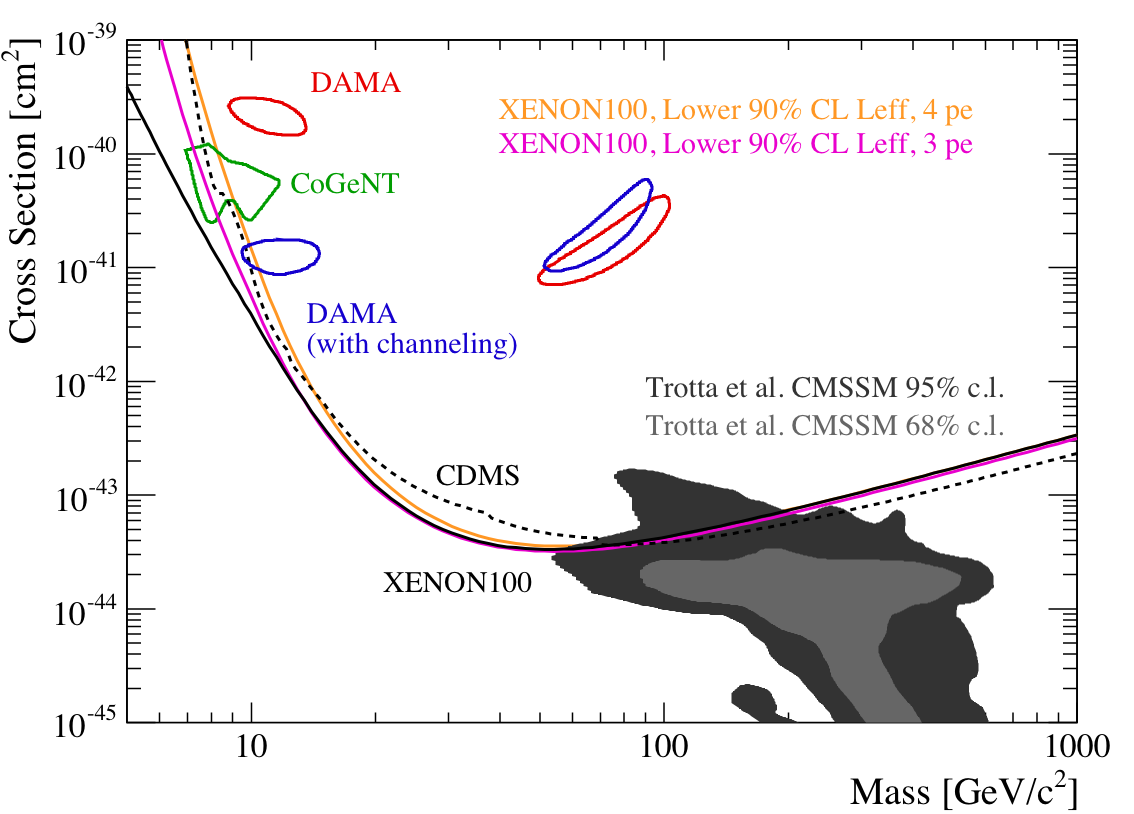}
\caption{90\% confidence limits for the global fit of ${\cal L}_{\text{eff}}$ with a threshold of 4\:PE (black), and curves for the 90\% lower contour of ${\cal L}_{\text{eff}}$ at thresholds of 4\:PE (yellow) and 3\:PE (magenta). We used the following astrophysical parameters: galactic escape velocity =  544 km/s,  WIMP density = 0.3 GeV/cm$^3$, solar velocity = 220 km/s.}
\label{fig:limits}
\end{figure}

Fig.\:\ref{fig:limits} shows the impact of the conservative choice of the 90\% lower contour of ${\cal L}_{\text{eff}}$ on the upper limits derived from the first XENON100 dataset, both for our pre-defined threshold of 4~PE and a lower threshold of 3~PE. In general, other uncertainties play a role at low WIMP masses as well, e.g., the galactic escape velocity. We used a value of 544 km/s \cite{ref:escape}. We further used a local WIMP density of 0.3 GeV/cm$^3$ and a solar velocity of 220 km/s, different (and less optimistic) from the values used in \cite{ref:collar}. Our limits are distinctly different from the curves presented in \cite{ref:collar}. 

In conclusion, we agree with the authors that the current situation on ${\cal L}_{\text{eff}}$ in LXe is far from optimal and must be clarified especially at the lowest Xe recoil energies. However, in our manuscript we have properly taken into account the uncertainty by using ${\cal L}_{\text{eff}}$ obtained from a global fit to all published direct measurements and by cross-checking the results with the lower 90\% CL contour together with a very conservative extrapolation for $E_r<5$~keV$_r$.

\end{document}